# Focus-Induced Photoresponse: a novel optoelectronic distance measurement technique

**Authors:** Oili Pekkola[1], Christoph Lungenschmied[1,*], Peter Fejes[1], Anke Handreck[1], Wilfried Hermes[1], Stephan Irle[2], Christian Lennartz[1], Christian Schildknecht[1], Peter Schillen[1], Patrick Schindler[1], Robert Send[1], Sebastian Valouch[1], Erwin Thiel[2], Ingmar Bruder[1]

**Affiliations:**

[1]trinamiX GmbH – a subsidiary of BASF SE, Industriestr. 35, 67063 Ludwigshafen, Germany.
[2]ERT Optik Dr. Thiel GmbH, Donnersbergweg 1, 67059 Ludwigshafen, Germany.
*Correspondence to:  christoph.lungenschmied@trinamix.de

**Abstract**: We present the Focus-Induced Photoresponse (FIP) technique, a novel approach to optical distance measurement. It takes advantage of a widely-observed phenomenon in photodetector devices: a nonlinear, irradiance-dependent photoresponse.  This means that the output from a sensor is dependent on the total number of photons incident and the size of the area in which they fall.  With a certain arrangement of sensor and lens, this phenomenon will cause the output of the sensor to change based on how far in or out of focus an object is.  We call this the FIP effect. Here we demonstrate how to use the FIP effect for distance measurements. We show that this technique works with different sensor materials, device types, as well as visible and near infrared light. In principle, any sensor exhibiting a photoresponse that depends nonlinearly on irradiance could be used with the FIP technique.  It is our belief that the FIP technique can become an important method for measuring distance.

**One Sentence Summary:** We introduce a novel distance measurement technique, which utilizes detectors with an irradiance-dependent photoresponse, demonstrate how to apply it to measurement challenges using two types of photodetectors in the visible and the IR regime, and model the irradiance dependence of these detectors.

**INTRODUCTION**

Optical distance measurement is already key to diverse applications throughout a wide range of industries. In the coming years, it is expected to gain even more importance due to the emergence of disruptive technologies such as machine vision and autonomous driving. These technologies have the power to revolutionize the world around us. However, in order to do so, further advances in optical depth sensing are required (*1, 2*).

Technologies like time-of-flight (ToF) and image-based depth sensing rely on advanced manufacturing techniques such as highly developed lithography processes and CMOS technologies (*3*). Since much of this development has been focused on silicon, it has become the de facto standard in optical sensors. However, the optical characteristics of silicon are somewhat limited. It is only sensitive to the visible and near-infrared (IR) region of light ranging from 350 – 1,100 nm (*4*). The optimal performance is limited to the visible range, whereas NIR absorption is rather weak. Improved sensitivity in the IR regime would offer significant advantages in terms of eye safety (*5*), night vision, and visibility in foggy conditions or through smoke (*6*). Materials that feature narrower absorption bands could make measurements more robust against stray light and background illumination. In many cases, even if adapting other materials than silicon to the manufacturing processes is technically possible, it may not be economically viable. If these processes were not required, materials and device types could be adopted to suit the sensing application, rather than the other way around.

Here we introduce Focus-Induced Photoresponse (FIP), a novel method to measure distances. In a FIP-based system, distance is determined by using the analog photoresponse of a single pixel sensor. This means that the advantages of high-density pixelation and high-speed response are not necessary or even relevant for the FIP technique. High resolution can be achieved without the limitations of pixel size, and detectors selected for a FIP system can be orders of magnitude slower than those required by ToF based ones. A system based on FIP does not require advanced sensor manufacturing processes to function, making adoption of unusual sensors more economically feasible.

**Irradiance-dependent photoresponse of photodetectors**

In the FIP technique, a light source is imaged onto the photodetector by a lens. The size of its image depends on the position of the detector with respect to the focused image plane. FIP exploits the nonlinearly irradiance-dependent photoresponse of semiconductor devices. This means that the signal of a photodetector not only depends on the incident radiant power, but also on its density on the sensor area, the irradiance. This phenomenon will cause the output of the detector to change when the same amount of light is focused or defocused on it. This is what we call the FIP effect.

In most reports on nonlinear photoresponse, a change in irradiance has been achieved by varying the incident radiant power. However, an irradiance variation over several orders of magnitude as well as irradiances of over 1 sun can also be realized by focusing the light in a smaller spot. Irradiance-dependence is a commonly-observed phenomenon among various photodetector technologies operating from the UV to the IR regime, making them suitable for the FIP technique. For many thin-film solar cell technologies, a nonlinear photoresponse at low irradiance levels has been reported. Trapping of charge carriers as well as photoconductivity have been identified to decrease the responsivity at low light levels (*7*, *8*). On the other hand, high light intensities exceeding 1 sun can reduce the current collection in solar cells, decreasing the photoresponse of these devices at high irradiances (*9*, *10*). The reduced photovoltaic performance has been attributed to a change in recombination mechanisms and a change in series resistance with irradiance (*9*). In organic solar cells, recombination has been found to depend on the charge carrier concentration (*11-13*). In an experimental setup similar to that of the FIP technique, a reduced photoresponse in thin-film solar cells as well as PbS and HgCdTe photoconductors has been reported when only a small area of the device receives high-intensity illumination (*10*, *14-16*).

## RESULTS

**The FIP effect in dye-sensitized solar cells**

In dye-sensitized solar cells (DSSC), the dependence of the photovoltaic performance on light intensity is published in detail (*17-19*). A nonlinear photoresponse to modulated light is reported for low light intensities. DSSC contain a mesoporous (mp) $TiO_2$ layer sensitized with dye molecules. mp-$TiO_2$ acts as the electron transporting material, and its pores are filled with a hole transporter. In the experiments presented below, the hole transporting layer is made of a solid film of the organic material spiro-MeOTAD (*20*); the device is hence referred to as a solid-state DSSC (sDSSC). Charge transport in the mp-$TiO_2$ structure is strongly impeded by localized states in the band gap. These traps are occupied by photogenerated electrons relaxing into these states. The electron diffusion coefficient in the mp-$TiO_2$ is found to increase with the electron concentration (*17, 18, 21, 22*). A higher density of absorbed photons will yield more and therefore faster electrons per unit area.

The effect can be seen in the photocurrent transients shown in Fig. 1a. The sDSSC sample was illuminated through a lens by a square wave modulated 530 nm LED. While the distance between the LED and the lens was kept constant, the spot diameter was varied between 0.1 mm² and 19 mm$^2$ by moving the cell along the optical axis around the focused image plane. Assuming a uniform light distribution, this corresponds to irradiances between 25 – 6,000 W/m² (for details about the spot size and irradiance calculations, see Supplementary Information S1 and S2). We observe that increasing the irradiance strongly shortens the rise time of the photocurrent. At the two highest irradiance levels shown in Fig. 1a, the identical equilibrium photocurrent is reached within the pulse period and both currents reduce to zero within the dark period. All measured transients at lower irradiance levels are already too slow to reach equilibrium or decay to zero within the pulse duration. We interpret this behavior as a direct consequence of the reported increase of the diffusion coefficient of the electrons in $TiO_2$ with irradiance.

Figure 1b shows the alternating photocurrent density of the sDSSC as a function of the irradiance at modulation frequencies between 75 Hz and 975 Hz. Irradiances between 10 and 10,000 W/m² are covered by moving the sDSSC behind the lens in 0.1 mm steps and thereby changing the size of the image on the sensor. The scatter points represent the photocurrent densities measured at each position. The areas of the light spots were calculated with paraxial optics and assumed to be uniformly illuminated (see Supplementary Information S1). To test the validity of this assumption, we also obtained beam profiles by raytracing and used a segmentation approach to parametrize the photocurrent densities as a function of the irradiance. The model is described in detail in Supplementary Information S3. The resulting photocurrent densities are plotted as lines in Fig. 1b. We find good agreement with the data obtained by assuming uniformly illuminated light spots. The dashed line with a slope of unity represents the linear detector response at high intensities and is added as a guide for the eye. At low modulation frequencies, the photocurrents are close to the steady-state level. With increasing frequency, the deviation from linearity at low irradiance levels increases. The alternating photocurrent induced by a modulated light source can thus be larger at higher photon densities even though the total amount of light on the sample stays constant: the solar cell works more efficiently at high irradiance. This is the signature of the FIP effect in sDSSC.

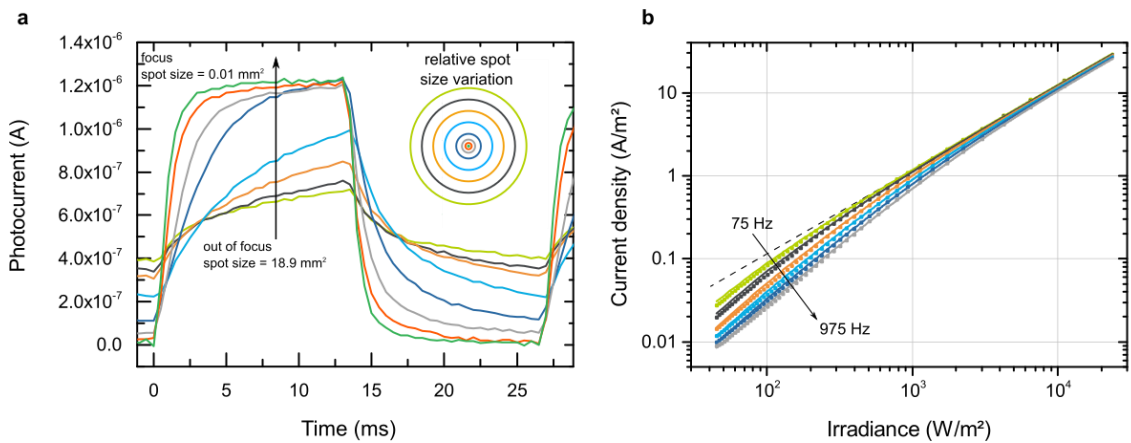

*Figure 1. Nonlinear photoresponse of sDSSC. a, Transient photocurrent response of a sDSSC to pulsed illumination through a lens. The device is placed at various positions on the optical axis, hence the same incident radiant power is distributed over different surface areas. The relative variation of the light spot sizes is illustrated by the circles. b, Alternating photocurrent of a sDSSC as a function of the irradiance for different modulation frequencies. The scatter points assume uniform illumination of the light spot. The solid lines are based on a raytracing model and the segmentation of the light spot. The dashed line with a slope of 1 acts as guide to the eye.*

**Measuring distances with the FIP technique**

We demonstrate how the FIP effect can be utilized for distance measurements with a setup schematically depicted in Fig. 2. Its components are a modulated LED[1], a consumer grade camera lens, two photodetectors, and a signal processing unit. The lens collects and focuses the light of the LED. The semitransparent sensors are placed behind the lens near its focal plane. The modulated light of the LED generates an alternating photoresponse, which is then amplified and recorded using lock-in or Fourier transform techniques. The spot size on the sensor changes with the distance between the light source and the lens as the position of the focused image plane shifts. The active area must therefore be large enough to accommodate the maximum spot size within the desired measurement range.

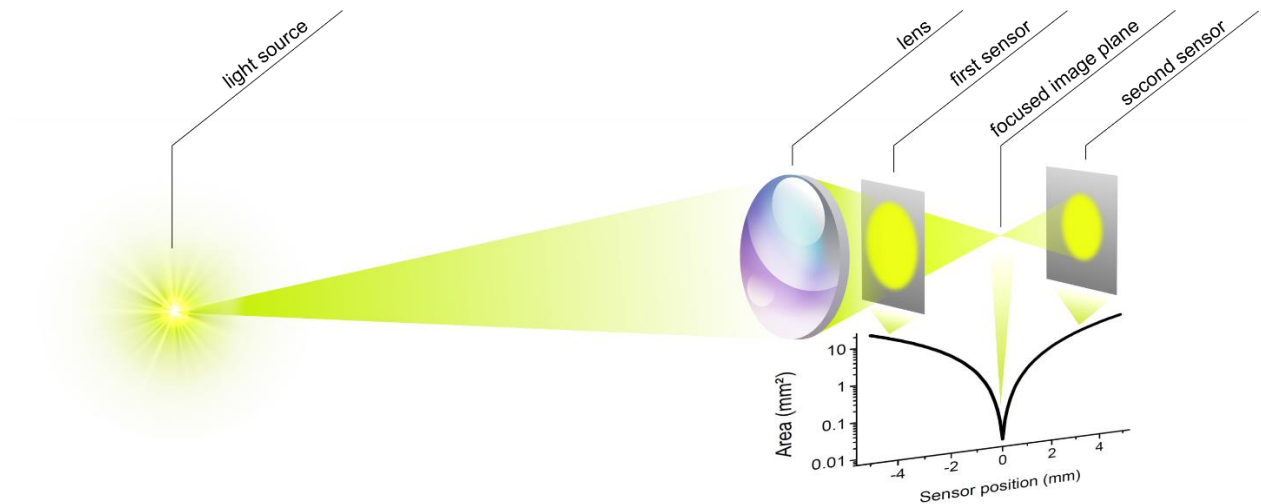

*Figure 2. A typical setup for measuring distances with the FIP technique. The size of the light spot and therefore the irradiance depends on the sensor position behind the lens. The graph shows the illuminated area around the position of the focused image plane for the experiments presented in Fig. 1.*

There are two main contributions to the sensor output. The first is due to the FIP effect: when the modulated LED is positioned at a certain distance from the lens, the measured photocurrent depends of the irradiance on the sensor, i.e. how well the light is focused. Secondly, the photocurrent is impacted by the total amount of light collected by the lens. When the LED is moved away from the lens, this contribution decreases. If the radiant power of the light source is unknown, the photoresponse of a single detector at any given position behind the lens does not allow for an unambiguous distance determination. Whether the LED is distant and bright or close and dim cannot be distinguished. To solve this problem, we use two detectors in the beam path. By calculating the ratio of the two photoresponses, the distance dependence as well as fluctuations in the output of the light source cancel out. Due to the FIP effect, the quotient changes with the distance, yielding a unique signature for each LED position.

---

[1] Instead of LEDS any light source, either actively emitting or reflecting light may be used. Alternatively to the camera lens, any optical element that captures the light and focuses it, such as lenses or mirrors, may be used.

**Demonstrating the FIP technique with sDSSC**

We have realized such a setup by using two semitransparent sDSSC as sensors and a modulated green LED. The results are summarized in Fig. 3. Figure 3a and 3b show the responsivity of the sDSSC as a function of their position behind the lens. The responsivity was calculated by normalizing the short circuit current to the total radiant power incident on the device. The sensor closer to the lens is referred to as the first sensor, and the one further from the lens as the second sensor. Due to the FIP effect, the maximum photocurrent is recorded when the sensor is positioned in the focused image plane. In this case, the LED light is focused on the sensor. A reduction in the current is observed as the light spot widens symmetrically. With increasing LED distance, the focused image plane moves towards the focal plane of the lens, leading to a shift of the photocurrent peaks. As shown in Fig. 3, we observe that the responsivity curves intersect at the focal plane of the lens. In this point, the irradiance in the light spot is independent of the LED distance, and only the diameter of the light spot changes (for the derivation, see Supplementary Information Section S4). This observation is consistent with our assumption that the FIP effect is induced by the irradiance-dependent sensor response.

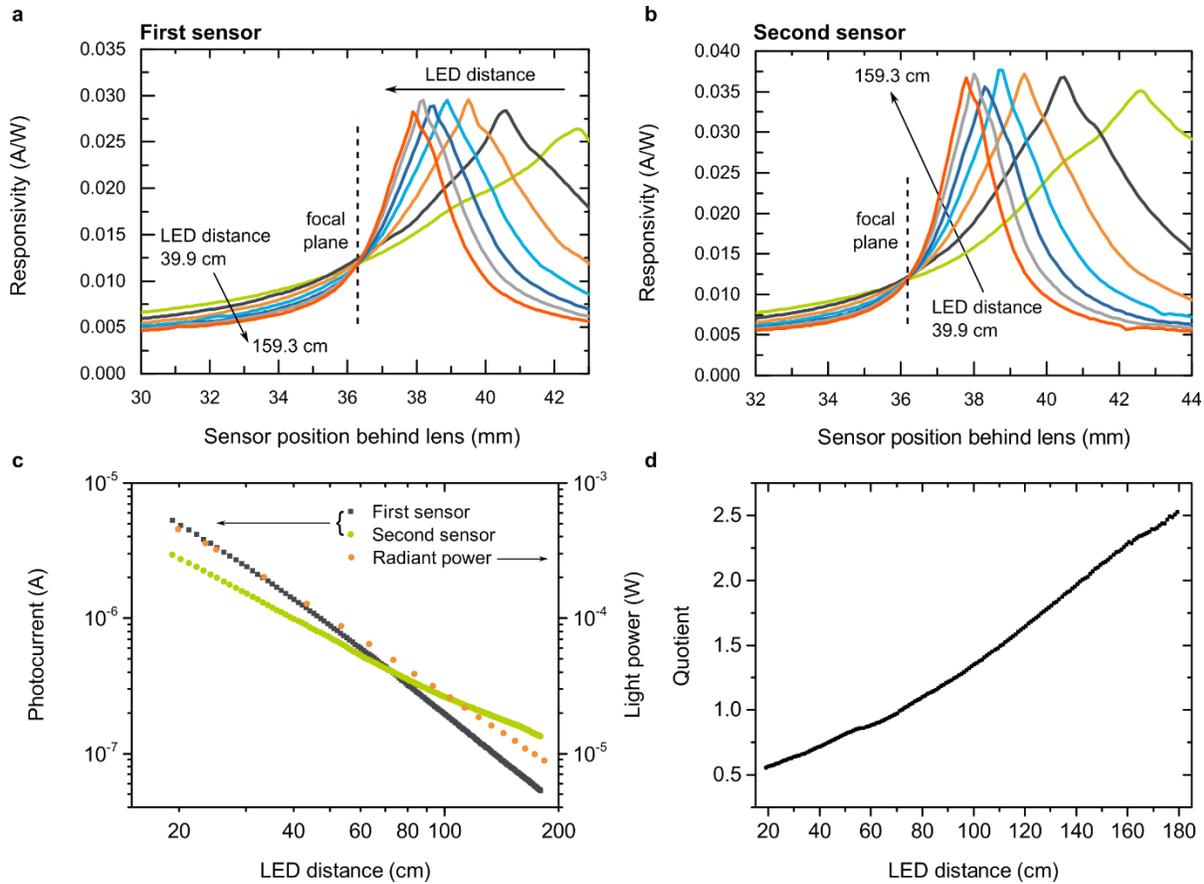

***Figure 3**. distance measurement with sDSSC. a,b, Responsivity of the first (a) and second (b) sDSSC sensor for a range of positions behind the lens at various LED distances. The second sensor is illuminated through the first. c, The absolute photocurrent of both sensors and the radiant power on the first sensor as a function of LED distance. d, The quotient of the photocurrents as a function of LED distance.*

The positions of the sensors in the beam path may be adjusted to the specific measurement problem. In the presented case, we measure distances in the range between 20 cm and 1.8 m (Fig. 3c and 3d). The first sensor is positioned 33.3 mm behind the lens, between the lens and its focal plane. The second sensor is placed 37.6 mm behind the lens. With increasing LED distance, the responsivity of the first sensor decreases, while it increases for the second sensor. Therefore, a large change in the ratio of their photocurrents is achieved over the measurement range. The absolute photocurrents of both sensors as well as the incident radiant power are plotted in Fig. 3c. The photocurrents are dominated by the inverse dependence of the radiant power on LED distance. The deviation from linearity with irradiance is visible as a difference in slopes. The resulting quotient increases monotonically over the entire measurement range (Fig. 3d), assigning a single value to each object distance. With this calibration curve, the LED distance can be directly determined by simply measuring the individual photocurrents of both cells and calculating their ratio.

**The FIP effect in PbS photoconductors**

The low-irradiance nonlinearity observed in DSSC at sufficiently large frequencies causes a maximum photoresponse when the device is placed in focus. In this situation, the irradiance and consequently the efficiency of the device are maximized. For other thin-film solar cell and photoconductor technologies, a reduced photoresponse has been reported when only a small area of the device receives high intensity illumination (*10, 14, 15*). We therefore expect the photoresponse of these devices to reach a minimum in focus, as the efficiency decreases with irradiance. PbS photoconductors used as IR sensors are an example of such behavior. In contrast to photovoltaic detectors, photoconductors do not generate a photocurrent. Instead, their resistance changes upon illumination. Even though the functionality differs fundamentally from DSSC, these devices can be used for distance measurements with the FIP technique. PbS detectors are opaque and therefore cannot be used in sensor stacks like the sDSSC presented above. Instead, a beam splitter may be used to position both sensors at appropriate distances from the lens.

Figure 4a shows the responsivity of a PbS detector to modulated 1,550 nm illumination as a function of its position behind the lens for various LED distances. The photoresponse is recorded as a voltage using an amplifier circuit. We determine the responsivity by normalizing the photoresponse to the radiant power reaching the detector. A minimum in the photoresponse is observed when the light spot is focused on the detector.

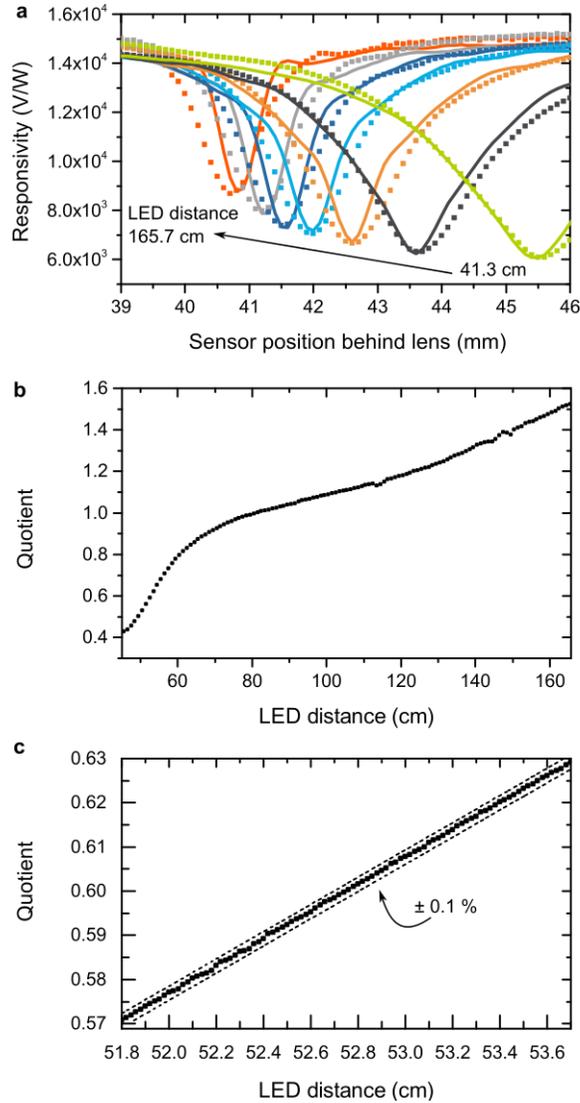

*Figure 4. Distance measurement with PbS. a, responsivity of a PbS photoconductor for a range of positions behind the lens at various LED distances. The device is modeled as a network of infinitesimally small photoconductors with a linear response to irradiance. The lines represent best fits to the experimental data. b, the quotient of the photoresponse of the PbS device at 40.4 mm and at 45 mm as a function of LED distance. c, detail of the quotient curve with an indication of a corridor of ±0.1 % of the distance to illustrate the obtained resolution.*

We have modeled the responsivity of the tested PbS photoconductor to light of various spot sizes and irradiance levels. The device is simulated as a two-dimensional grid of connected infinitesimally small photoconductor elements. By performing a limiting process, we can model the grid with an elliptic partial differential equation, i.e. $\nabla \cdot \left(\frac{\nabla \varphi}{R(x,y)}\right) = \nabla \cdot (\sigma(x,y)\nabla \varphi) = 0$, where

φ(x,y) is the electric potential, R(x,y) the local resistance and σ(x,y) the conductance (more details in Supplementary Information S5). This model can be solved using the Finite Elements Method. The dark conductance $\sigma_d$ of every point of the PbS photoconductor is modified by illumination. The conductance of an illuminated point is described as $\sigma = \sigma_d + \Delta\sigma$. $\Delta\sigma$ is assumed to be directly proportional to the irradiance ($E \sim \frac{W}{m^2}$), i.e. $\Delta\sigma = p \cdot E$. The irradiance is calculated using paraxial optics and assumes uniformly illuminated light spots (see S1). We have fitted the measured data plotted in Fig. 6a using $\Delta\sigma = 0.298 \frac{mm}{V^2} \cdot E$.

The resulting curves are shown as lines in Fig. 4a. We interpret the good agreement between the measured data and our model as a confirmation of the assumed linear dependence of Δσ on irradiance for the tested PbS photoconductor within the studied irradiance regime from 0.3 W/m² to 5,000 W/m². Even if a photoconductor reacts perfectly linearly to the irradiance, a FIP effect is observed when the active area is only partially illuminated. This behavior is consistent with experimental data (*15*) and shown formally in Supplementary Information S5.

The FIP effect in PbS photoconductors is used for distance measurements by assigning a photoresponse quotient to measured distances as shown in Fig. 4b. The ratio of the photoresponses was determined for sensor positions at 45 mm and 40.4 mm behind the lens. The resulting quotient increases over the studied measurement range, enabling accurate distance measurements between 45 – 165 cm. Figure 4c depicts a small range to illustrate the resolution of distance measurements with the FIP technique. LED positions as close together as 500 µm can be distinguished by the photoresponse quotient at a distance of 52 cm, corresponding to a depth resolution of better than 0.1 %.

**CONCLUSIONS AND OUTLOOK**

We have demonstrated that the FIP technique is a new and versatile method for measuring distance. The differences between it and traditional methods like ToF and triangulation open the door for different types of measurements. FIP sensors can be extremely simple. They do not need to be arrayed or operate at high speed. Many applications could benefit from using wavelengths outside the bounds of traditional distance measurement sensors; with FIP this is possible. The only requirement is that the sensor displays the FIP effect, which many materials do. We have observed the FIP effect in various thin-film photovoltaic device technologies such as DSSC, amorphous silicon, CdTe, CIGS, CIS, CZTS, as well as in organic solar cells (*23*) and PbS photoconductors. The samples we tested were either purchased or produced using standard techniques. In this article, we have shown resolution of below 500 µm at a distance of 50 cm. In our supplementary information, distance measurements up to 70 m can be found (Section S6). We believe that research and device optimization will further improve these results.

The technique of FIP can be combined with other technologies, to create systems with even more functionality. A device sensitive to the x, y and z coordinates (*24*) of a light spot can be created by using commercially available position sensitive devices (PSD) as the sensors. Simultaneous tracking of multiple light spots is possible if they have different modulation frequencies. It is also possible to utilize projected light spots instead of actively emitting ones. This allows lasers to be used as the light sources instead of the LEDs we have presented here. An advantage of this method is that the position of the laser's origin does not impact the measurement. The FIP technique only measures the distance to the light spot. With improved understanding and further development of the technique, FIP can become an important distance measurement technique.

Acknowledgments: We acknowledge the help of John Dowell and Ines Kühn in improving the overall quality of the manuscript. We wish to thank Peter Haring Bolivar from the University of Siegen for discussions and the scientific support, as well as Peter Erk, Karl Hahn, and Harald Lauke from BASF SE for their continued support.

**Methods**

*Fabrication of sDSSC*. The FTO substrates (Pilkington glass) were first cleaned with a glass detergent, then rinsed with water and cleaned with acetone and isopropanol. Subsequently, the substrates were ozone treated for 30 min (Novascan PSD Series Digital UV Ozone System). After that, a $TiO_2$ blocking layer was deposited on the substrates via spray pyrolysis. 9.72 g titanium diisopropoxide bis(acetylacetonate) was dissolved in 100 ml ethanol. 25 spray cycles were performed at 350 °C. After the pyrolysis, the samples were annealed at 350 °C for 30 min. For the deposition of a mesoporous TiO2 layer, transparent titania paste (Dyesol, average particle size 20 nm) was mixed with ethanol in a ratio of 1:3. The solution was spin coated at 3700 RPM for 30 s, and the films were sintered subsequently at 450 °C for 30 min. The TiO2 films were immersed in a 5 mM dye (N-Carboxymethyl-9-(7-(bis(9,9-dimethyl-fluoren-2-yl)amino)-9,9-dihexyl-fluoren-2-yl)perylene-3,4-dicarboximide) solution in toluene for 1 h. After that the samples were rinsed with water and dried with nitrogen. 100 mg/ml hole conductor 2,2',7,7'-Tetrakis[N,N-di(4-methoxyphenyl)amino]-9,9'-spirobifluorene (Spiro-MeOTAD) in chlorobenzene was mixed with 20 mM bis(trifluoromethane) sulfonamide lithium salt in cyclohexanone and 2.5 mg/ml vanadium pentoxide, and oxidized in air for 1 h. Subsequently, vanadium pentoxide was removed by filtering the solution through a 0.2 µm PTFE filter. The solution was spin coated at 2000 RPM for 30 s and the samples were left to dry for 30 min. We used PEDOT:PSS (Clevios F HC Solar) as the counter electrode. The dispersion was filtered with a 0.45 µm PTFE filter and spin coated at 2000 RPM for 30 s. After that, the samples were dried on a hot plate at 90 °C. Finally, 200 nm Ag contacts were evaporated on top of the PEDOT:PSS film using a custom-made Creavac thermal evaporator.

*Transient photocurrent*. The sDSSC was illuminated with a 530 nm LED (Thorlabs M530L3). The LED was modulated at 375 Hz with square wave pulses and a duty cycle of 50 %. The light was focused with an aspheric lens (Thorlabs AL2520-A) that was positioned at 35.2 mm from the LED. The sensor was mounted on a translational stage that allowed movements along the optical axis. The light power on the sensor was 465 µW, and the size of the image on the sensor was changed by moving the sensor on the optical axis. The transient photocurrent was amplified with a Femto DLPCA transimpedance amplifier (gain 104 V/A) and recorded with a National Instruments PXIe-4492 measurement card.

*Photocurrent as a function of photon density and modulation frequency.* The measurement setup was identical to that of transient photocurrent as described above. The alternating photocurrent was amplified with a Femto DLPCA transimpedance amplifier (gain 104 V/A) and recorded with a Behringer U-Phoria UMC202HD sound card. The LED was modulated at 75, 175, 375, 575, 775, and 975 Hz, and the light power on the sensor was 855 µW.

*FIP distance measurements with DSSC.* A stack of two semitransparent sDSSC was illuminated with a square wave pulsed LED (Thorlabs M530L3) at 530 nm. The LED was modulated with square wave pulses at 475 Hz. The light was focused with a Nikkor 50 mm f/1.2 lens. The distance between the sensors in the stack was 4.3 mm. The stack was mounted on a translational stage and the LED on a rail that allowed movements along the optical axis. The radiant power on the first sensor at LED distance of 13.5 cm was 790 µW. The photocurrents were amplified with two Femto DLPCA-200 transimpedance amplifiers (gain $10^4$ V/A) and recorded with two lock-ins (SR 850, Stanford Research Systems).

*FIP distance measurements with PbS and the photoconductor model.* A commercial PbS photoconductor (Hertzstück$^{TM}$, active area 1 cm x 1cm) was illuminated with an LED at 1,550 nm (Thorlabs M1550L3). The LED was modulated with square wave pulses at 606 Hz and its light was focused with a Nikkor 50 mm f/1.2 lens. The photoresponse was measured using a voltage divider including a 2 MΩ resistor to match the dark resistance of the PbS device of similar resistance. A voltage of 100 V was applied to the photoconductor and the 2 MΩ resistor, hence an electric field of 50 V/cm was present across the active area of the photodetector. The photoresponse to the modulated LED was then determined with a Behringer U-Phoria UMC202HD sound card connected via a unity gain buffer. The experimental data are obtained using an FFT with a bandwith of 1 Hz. This can be interpreted as a smoothing process. In order to remove numerical perturbations, the simulated results were smoothed by using a moving average filter as well. The

PbS photoconductor was mounted on a translational stage and the LED on a rail that allowed their movements along the optical axis. The radiant power on the sensor at an LED distance of 13.5 cm was 35.1 µW.

**Supplementary information**

**S1 – Modelling an optical image with paraxial optics**

We use paraxial optics to estimate the size of an optical image on the FIP sensor. The optical setup is described by the paraxial approximation, i.e. the model is based on the thin lens equation

$$\frac{1}{f} = \frac{1}{z} + \frac{1}{b} \qquad (S1-1)$$

where $f$ is the focal length of the lens, $z$ the distance between the light source and the lens and $b$ the distance between the lens and the focused image of the light source. It should be noted that the model is an approximation. The irradiance distribution within the image or the properties of the lens are not considered.

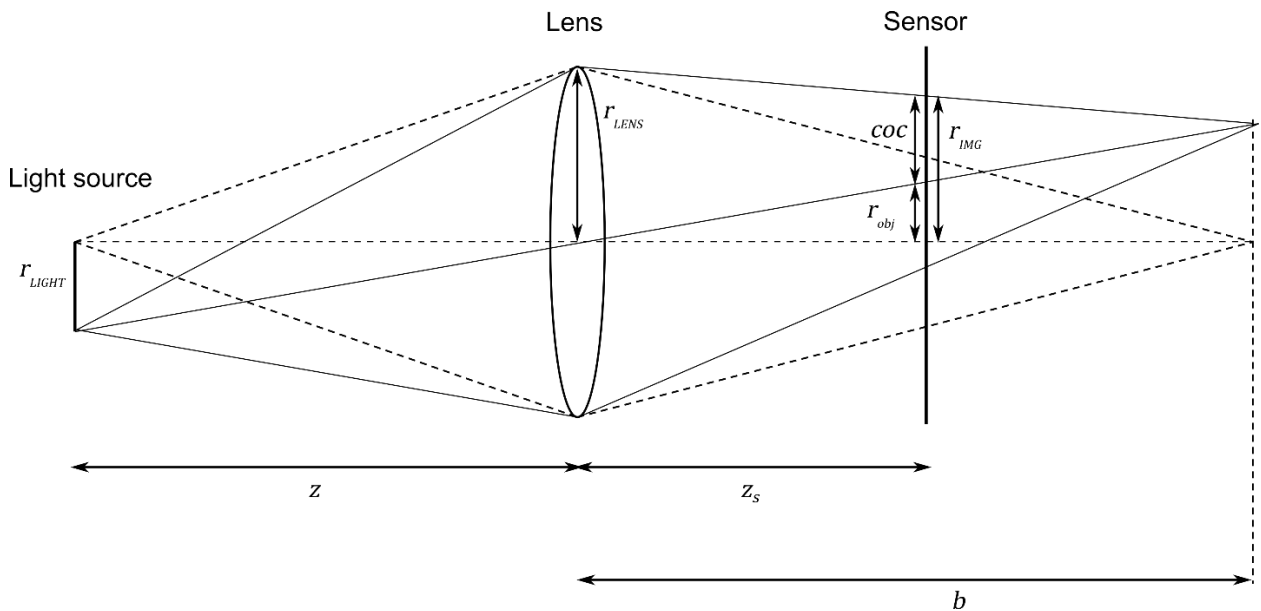

*Fig. S1.1    Optical setup*

The disk-shaped light source has a radius $r_{LIGHT}$. The sensor is placed at a distance $z_s$ behind the lens. The image of the light source on the sensor has a radius $r_{IMG}$. It consists of the image of the light source without blur, $r_{OBJ}$, and the circle of confusion $coc$:

$$r_{IMG} = r_{OBJ} + |coc| \qquad (S1-2)$$

The circle of confusion is determined by the intercept theorem, i.e.

$$\frac{coc}{b - z_s} = \frac{r_{LENS}}{b}. \qquad (S1-3)$$

Combining equations $S1-1$ and $S1-3$, the circle of confusion yields to

$$coc = r_{LENS}\left(1 - \frac{z_s}{b}\right) = r_{LENS}\left(1 - z_s \cdot \left(\frac{1}{f} - \frac{1}{z}\right)\right) = r_{LENS}\left(1 - z_s \cdot \left(\frac{z-f}{fz}\right)\right). \qquad (S1-4)$$

The object size on the image $r_{OBJ}$ is also determined by the intercept theorem, i.e.

$$\frac{r_{OBJ}}{z_s} = \frac{r_{LIGHT}}{z}. \qquad (S1-5)$$

The radius of the optical image $r_{IMG}$ is thus

$$r_{IMG} = r_{LENS}\left|1 - z_s \cdot \left(\frac{z-f}{fz}\right)\right| + \frac{r_{LIGHT}}{z} z_s. \qquad (S1-6)$$

This formula can be extended easily by using the diameter $d_{LENS}$ instead of the radius:

$$d_{IMG} = d_{LENS}\left|1 - z_s \cdot \left(\frac{z-f}{fz}\right)\right| + \frac{d_{LIGHT}}{z} z_s. \qquad (S1-7)$$

## S2 – Image sizes and photon densities in transient photocurrent measurements (Fig. 1a)

Distance of the sensor from the focused image plane: 0.2 – 5 mm, power of the light source: 465 µW

Sizes of the light spot were calculated with paraxial optics (Eq. S1-7) with following parameters:
- Distance of the light source   35.2 mm

- Diameter of the light source  2 mm
- Focal length of the lens      20 mm
- Working F#                    0.979877

| Distance from focused image plane (mm) | Spot area (mm²) | Irradiance (W/m²) |
|---|---|---|
| 0.2 | 0.08 | 6087.73 |
| 0.5 | 0.28 | 1650.56 |
| 1 | 0.91 | 510.09 |
| 2 | 3.25 | 143.08 |
| 3 | 66.18 | 66.18 |
| 4 | 12.24 | 37.99 |
| 5 | 18.90 | 24.61 |

## S3 – Simulation of beam profiles with ray tracing

Although paraxial optics allows for a qualitative understanding of the imaging process, a ray tracing model accounting for actual lens systems is needed. Since the sensor signal depends on irradiance, special care must be taken to account for the spatial distribution of photons over the sensor area. It is not possible to assign a single irradiance value to a specific beam profile in a FIP measurement, since each profile on the sensor consists of a characteristic distribution of local irradiances. To deduce the specific relation between irradiance and the current density of a certain sensor type, we have chosen the following ansatz:

We discretize the sensor area with a rectangular grid. At each sensor position, every pixel is assumed to behave as a local sensor that is exposed to a discrete irradiance $E_{i,s}$. The discrete irradiance depends on the radiant power distribution $\Phi_{local}$ at that pixel and sensor position:

$$E_{i,s} = \frac{\Phi_{local}(x_i, y_i, z_s)}{A_{local}}, \quad (S3-1)$$

where $x_i$ and $y_i$ are the pixel coordinates, $z_s$ the sensor position with respect to the lens and $A_{local}$ the area of each pixel. We now define the local response function $f_{local}$ as

$$f_{local} = p_1 \cdot E_{i,s} - p_2 \cdot E_{i,s} \cdot \exp(-p_3 \cdot E_{i,s}^{p_4}), \quad (S3-2)$$

where $p_1 - p_4$ are simulation parameters. The local response function gives the local current $I_{i,s}$ originating from the local irradiance:

$$I_{i,s} = f_{local}(E_{i,s}) \qquad (S3-3)$$

By summing over all local currents, we obtain the overall sensor response current $I_s$ for each sensor position $z_s$:

$$I_s = \sum_i f_{local}(E_{i,s}) \qquad (S3-4)$$

To obtain the parameters of the local response function, we used the simulated irradiation profiles at different sensor positions in combination with the experimentally measured photocurrents and performed a least-squares fit to obtain the best matching local response function.

Specifically, we used the following parameters:

- Lens: Thorlabs AL2520M-A, Mounted Asphere, Ø25.0mm, EFL = 20.0mm, NA=0.54, -A Coating
- LED position with respect to the lens: 32.5 cm
- Wavelength: 530 nm
- Width of the discretized sensor: 7 mm
- Number of pixels per line/column: 1,001
- Number of simulation rays: 5,000,000

Parameters for different modulation frequencies:

| Frequency | p₁ | p₂ | p₃ | p₄ |
| --- | --- | --- | --- | --- |
| 975 Hz | 0.0010774475772490 | 0.0010906322277830 | 0.026971714492125 | 0.531302606764043 |
| 775 Hz | 0.001112257999756 | 0.001151022565467 | 0.034556007165795 | 0.523592469700944 |
| 575 Hz | 0.001149372503804 | 0.001279720508382 | 0.058938000077514 | 0.479743037095579 |
| 375 Hz | 0.001181978599514 | 0.001493075487758 | 0.109307540253059 | 0.430996445237502 |
| 175 Hz | 0.001221120774188 | 0.002960866589844 | 0.477235010400842 | 0.281457998864533 |
| 75 Hz | 0.001242651802542 | 0.012728796621985 | 1.658609785709937 | 0.166377333205729 |

## S4 - Modelling the irradiance of a photodetector through a lens

**Assumptions**:

- The light source is infinitesimally small (point light source) and emits uniformly in all directions
- The optical setup is described by the paraxial approximation, observing the thin lens equation S1-1
- The sensor is larger than the image of the light source

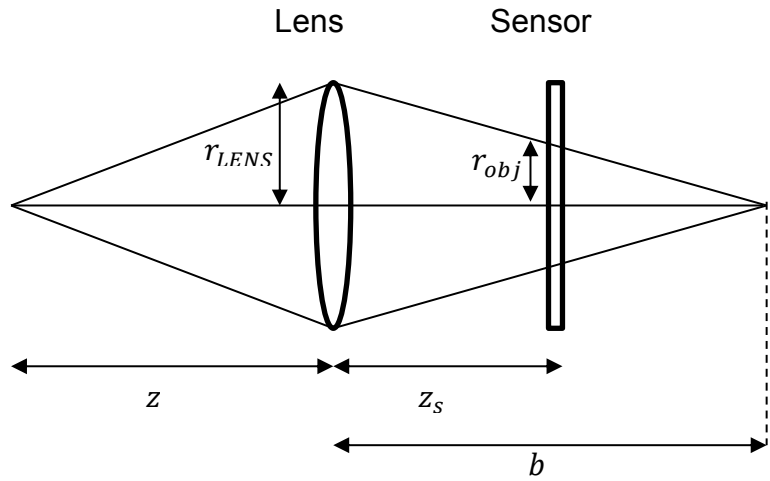

*Figure S4.1: Optical Setup*

In consideration of these assumptions, the optical image on the sensor is a circular disk. Its radius ($r_{obj}$) is given by

$$r_{obj} = \left| r_{LENS} \left(1 - z_s \frac{z-f}{zf}\right) \right|, \qquad (S4-1)$$

where $r_{LENS}$ is the radius of the lens and $z_s$ is the distance between lens and sensor.
The amount of light shining on the sensor is modeled by the radiant power $L$. For sufficiently large $z$, it decreases with the square of the distance between light source and lens. The parameter $\lambda_0$ characterizes the emitted light and the transmission properties of the lens.

$$L(z) = \frac{\lambda_0}{z^2} \qquad (S4-2)$$

The irradiance $E$ of the sensor is given by the distribution of the radiant power on the sensor

$$E(z) = \begin{cases} \dfrac{1}{\pi r_{obj}^2} L(z) & \|x\| \leq r_{obj} \\ 0 & \|x\| > r_{obj} \end{cases} \qquad (S4-3)$$

The nonlinear sensor response function to the irradiance is defined by $F$. The overall sensor response is the spatial integral over the irradiance, i.e.

$$I(z) = \int F(E(x)) dx = \pi r_{obj}^2 \cdot F\left(\dfrac{1}{\pi r_{obj}^2} \cdot L(z)\right). \qquad (S4-4)$$

The normalized sensor response $I_{norm}$ is defined by

$$I_{norm} = \dfrac{I(z)}{L(z)}. \qquad (S4-5)$$

**Iso-FIP theorem:**

Let the sensor position set to $z_s = f$. Then the following result is valid: For any sensor response function $F$ the normalized sensor response does not depend on the distance $z$.

**Proof**:

Let $z_s = f$, the radius of the illuminated disk reduces to

$$r_{obj} = \dfrac{r_{LENS} \cdot f}{z}. \qquad (S4-6)$$

Then the normalized sensor response $I_{norm}$ yields to

$$I_{norm} = \dfrac{\pi r_{LENS}^2 f^2}{z^2} F\left(\dfrac{z^2}{\pi r_{LENS}^2 f^2} L(z)\right) L(z)^{-1}. \qquad (S4-7)$$

Plug in the irradiance function on the sensor $L$:

$$I_{norm} = \dfrac{\pi r_{LENS}^2 f^2}{z^2} F\left(\dfrac{z^2}{\pi r_{LENS}^2 f^2} \dfrac{\lambda_0}{z^2}\right) \dfrac{z^2}{\lambda_0} = \dfrac{\pi r_{LENS}^2 f^2}{\lambda_0} F\left(\dfrac{\lambda_0}{\pi r_{LENS}^2 f^2}\right). \qquad (S4-8)$$

For the normalized sensor response, $z$ cancels out, hence the response does not depend on the distance $z$.

□

**Explanation:**

When the distance between light source and sensor ($z$) increases, the amount of light impinging on the sensor decreases. At the same time, the size of the optical image decreases. Both trends contribute to the amount of light per unit area, the irradiance of the sensor.

Assuming the characteristics of a point source and the validity of the thin lens approximation, we show that if the sensor is positioned in the focal plane of the lens (at $z_s = f$), the area of the optical image is inversely proportional to the square of the distance between light source and sensor. This yields that the reduction in the amount of light on the sensor cancels out the reduction in image size. Hence the irradiance of the image remains constant when $z$ is varied.

We assume that the quantum efficiency, as expressed by the sensor response function $F$, depends on the irradiance. Under the above assumptions, the irradiance is constant over of area of the image. Thus, the normalized sensor response of the detector positioned in the focal plane is identical for any distance $z$.

Even though the conditions in the actual experiments deviate significantly from the assumptions defined above, we find that responsivity curves at various distances for sDSSC (Fig. 3a), PbS photoconductors (Fig. 4a) and amorphous silicon (Fig. S6.1) intersect close to the focal length of the used lens.

## S5 - Modelling the photoresponse of partially illuminated photoconductors

We model a photoconductor device as an infinitesimally fine network of light dependent resistors. The current flow is governed by the Ohm's law, i.e. $U = R \cdot I$.

The goal is to derive a continuous model of the network by a limit process for mesh sizes $\Delta x$ approaching zero (Fig. S5.1). The result of the limit process is an elliptic partial differential equation that describes the current flow within a photoconductor.

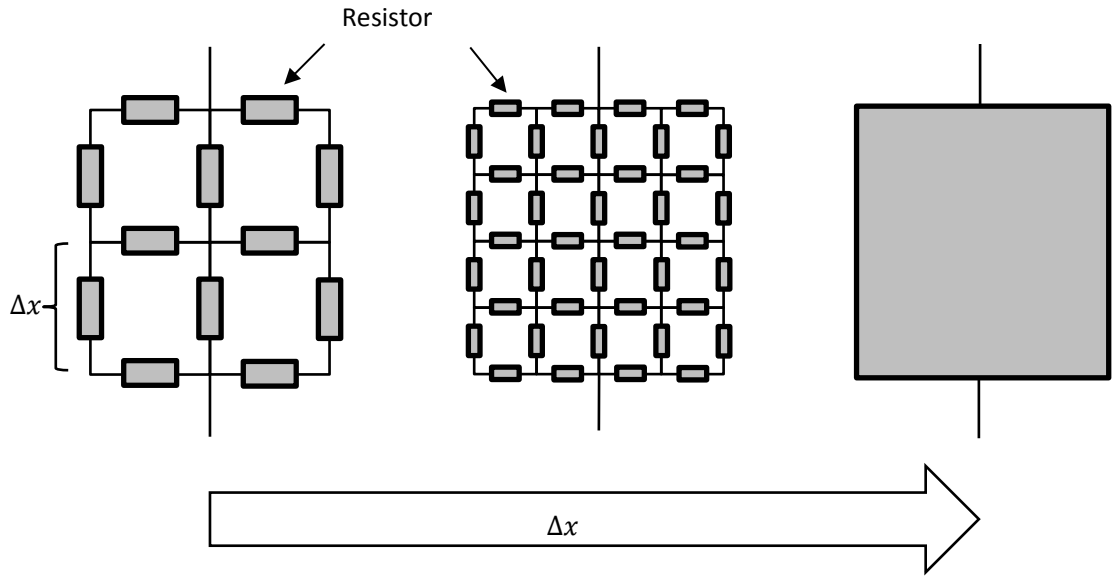

*Fig. S5.1    Visualization of a limit process of electricity network. The photoconductor is modeled as a coarse electricity network.*

The continuous model can be derived as follows:

We can interpret the solution of the resistor network as the solution of a two-dimensional finite-volume scheme. Each arc of the network should have the same length $\Delta x$. We assume that the values of a node correspond the value of a cell (see Fig. S5.2). For more details see: *Peter Schillen - Modelling and Control of Balance Laws with Applications to Networks – Dr. Hut - ISBN* 9783843922159 *– Section 3.6*. The voltage is given by the finite difference of the potential $\varphi$, e.g. $U_N = (\varphi_{i,j+1} - \varphi_{i,j})$.

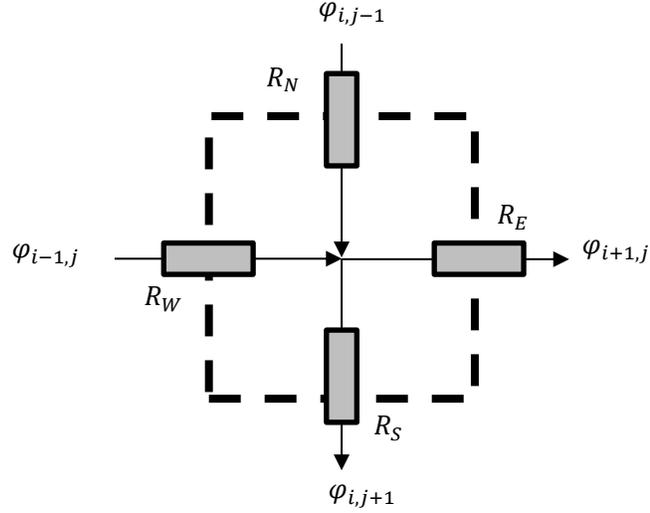

*Fig. S5.2    Scheme of a node in the resistor network*

Kirchhoff's current laws and Ohm's law yield the following equation

$$0 = \frac{I_S - I_N}{\Delta x} + \frac{I_E - I_W}{\Delta x} = \frac{(\varphi_{i,j+1} - \varphi_{i,j})R_N - (\varphi_{i,j} - \varphi_{i,j-1})R_S}{\Delta x^2 R_S R_N} + \frac{(\varphi_{i,j+1} - \varphi_{i,j})R_W - (\varphi_{i,j} - \varphi_{i,j-1})R_E}{\Delta x^2 R_E R_W} \quad (S5-1)$$

For $\Delta x \to 0$ the finite differences coincide with its derivatives.

$$0 = \frac{\partial_{xx}\varphi R(x,y) - \partial_x \varphi \partial_x R(x,y)}{R(x,y)^2} + \frac{\partial_{yy}\varphi R(x,y) - \partial_y \varphi \partial_y R(x,y)}{R(x,y)^2} \quad (S5-2)$$

It is necessary to assume that $R$ is weakly differentiable.

Finally, the limit yields a continuous model based on a partial differential equation

$$\nabla \cdot \left( \frac{\nabla \varphi}{R(x,y)} \right) = 0, \quad (S5-3)$$

for all $x \in \Omega$, where $\Omega$ is the cell domain. Note that Kirchhoff's voltage law is already fulfilled by the fundamental theorem of calculus.

Additionally, boundary conditions are required for unique solutions: The first case is that the boundary of a cell is connected to a voltage source. That means that the potential on the boundaries is described by a known function $\varphi_0$, i.e.

$$\varphi(x) = \varphi_0(x) \; \forall x \in \partial \Omega. \quad (S5-4)$$

The boundary of the cell domain $\Omega$ is denoted by $\partial \Omega$.

The second case is that there is no electrical connection between the cell boundary and a voltage source, e.g. isolation by air. Then the boundary condition is given by

$$\nabla \varphi(x) \cdot n = 0 \; \forall x \in \partial\Omega, \tag{S5-5}$$

where $n$ is the outgoing normal of the domain $\Omega$. Note that these two cases of boundary conditions can be mixed. An example is given in Fig. S5.3. The cell has two electrodes with a voltage of $U_0$ and two isolated edges.

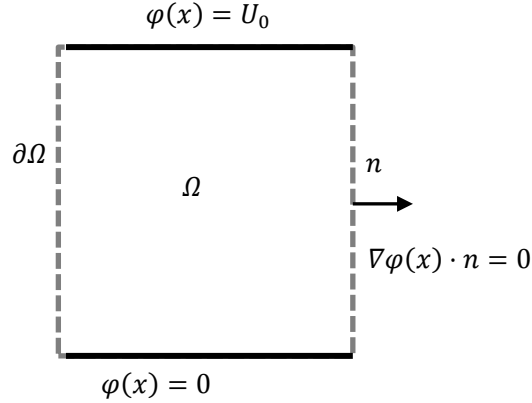

*Fig. S5.3    Example of a cell domain $\Omega$. The continuous edge (top and bottom) corresponds to the cell electrodes with a voltage of $U_0$. The dashed line (left and right) indicates the part without any voltage source.*

### Analytical solution for the one-dimensional case:

For the one-dimensional case $\partial_x \left( \frac{\partial_x \varphi}{R(x)} \right) = 0$ we can solve the equation analytically. First, we express the resistance in the form of conductivity, i.e. $R(x) = \frac{1}{\sigma(x)}$. By integrating over $x$ we get:

$$\varphi(x) = c_0 \int_0^x R(s) \, ds + c_1 \tag{S5-6}$$

We consider an area of length $b$, with boundary values $\varphi(0) = 0, \varphi(b) = U_0$, hence yielding a voltage of $U_0$. With the boundary values, it follows that, $c_1 = 0$, $c_0 = U_0 \frac{1}{\int_0^b R(s) ds}$. As one can see, $c_0$ is the current of the system. We assume that the conductivity of a photoconductor device changes with the light intensity and the geometry of the cell.

We consider a one-dimensional photoconductor. If the length is fixed, the conductivity only depends on the light intensity. The dark conductivity $\sigma_d$ changes upon illumination by $\Delta \sigma = p \cdot$

$E$, $\Delta\sigma$ hence depending linearly on the irradiance $E$. In Fig. S6.4 a one-dimensional photoconductor is shown schematically.

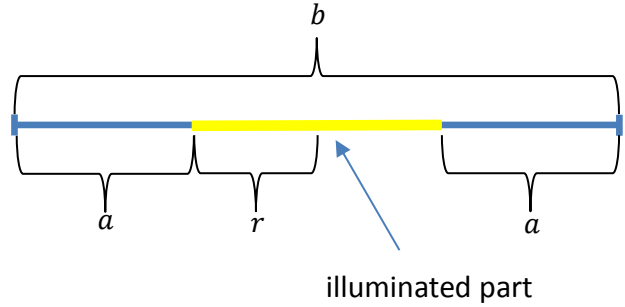

*Fig. S5.4*   *scheme of a partially illuminated one-dimensional photoconductor.*

Now we can compute $\int_0^b R(s)\,ds = \int_0^b \frac{1}{\sigma(s)}\,ds = \int_0^a \frac{1}{\sigma_d}\,ds + \int_a^{a+2r} \frac{1}{\sigma_d+pE}\,ds + \int_{a+2r}^{2(a+r)} \frac{1}{\sigma_d}\,ds =$

$2r\frac{1}{\sigma_d+pE} + 2a\frac{1}{\sigma_d} = \frac{2r\sigma_d+2a(\sigma_d+pE)}{\sigma_d(\sigma_d+pE)}$. Now we have $c_0 = \frac{\sigma_d(\sigma_d+pE)}{2r\sigma_d+2a(\sigma_d+pE)}U_0$.

**Theorem:** $c_0(E)$ is affine linear for $a = 0$ (i.e. the sensor is fully and uniformly illuminated) and non-linear for $a > 0$.

*Proo*f: We derive $c_0(E) = \frac{\sigma_d(\sigma_d+pE)}{2r\sigma_d+2a(\sigma_d+pE)}U_0$ w.r.t. $E$ and get $\frac{d}{dE}c_0(E) = \frac{p\sigma_d^2 r}{2(a(pE+\sigma_d)+\sigma_d r)^2}U_0$.

Trivially $\frac{d}{dE}c_0(E)$ does not depend on $E$ for $a = 0$, otherwise for $a \neq 0$ it does. This proves that $c_0(E)$ is affine linear for $a = 0$.

q.e.d.

**S6 – Long-range measurements over 72 m using amorphous silicon-based FIP detectors**

Long-range measurements up to 72 m are performed using an amorphous silicon based FIP detector. The distance to a modulated LED emitting at 660 nm[2] is determined using the FIP technique. The LED light is collected by a 95 mm diameter commercial lens[3] which directs the converging beam towards the sensors. A stack of two semitransparent amorphous silicon solar cells[4] operating at short circuit is used as detector. To cover distances beyond 36 m, a 20 x 30 cm

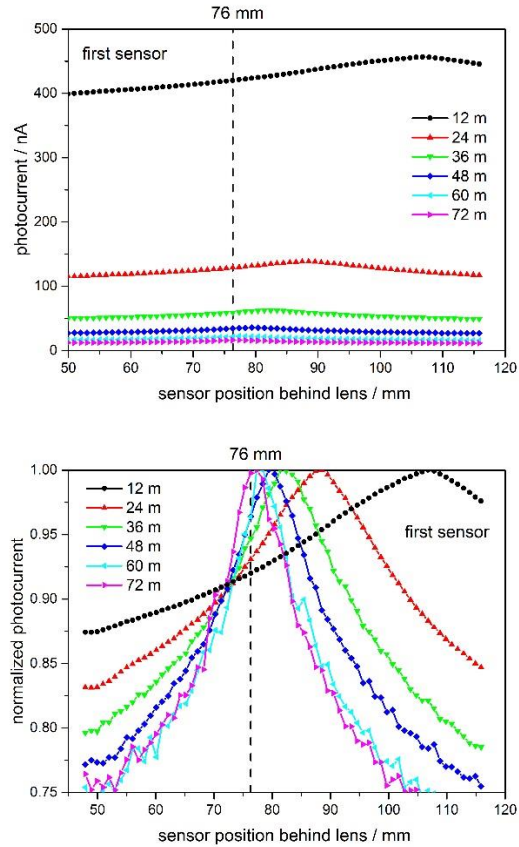

large mirror is used as reflector, allowing for measurements up to 72 m.

---

[2] Thorlabs model M660L3. 700 mW emitted optical power. 50% duty cycle, 2225 Hz, https://www.thorlabs.com/thorproduct.cfm?partnumber=M660L3

[3] Walimex Pro 500mm lens (http://www.walimexpro.de/en/video/video-lens/c-mount/produkt/walimex-pro-50063-dslr-mirror-c-mount-white.html)

[4] 40 x 40 mm active area, 500 nm i-layer thickness. Custom-made by Solems S. A.

*Fig. S6.1    Short circuit photocurrent of the first sensor at various positions behind the lens for several LED distances ranging from 12 m to 72 m. The dashed lines indicate the sensor position used for the quotient determination.*

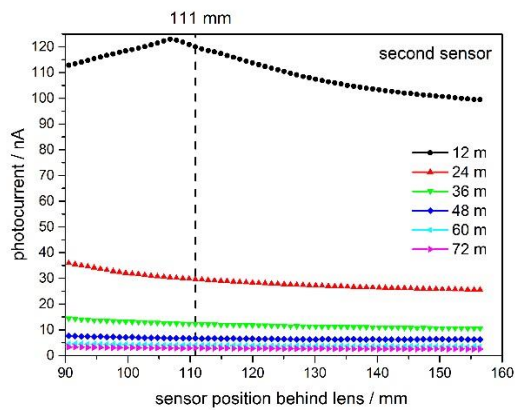

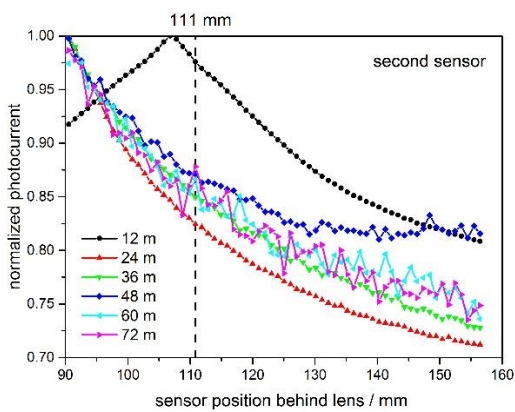

*Fig. S6.2    Short circuit photocurrent of the second sensor at various positions behind the lens for several LED distances ranging from 12 m to 72 m. The light passes through the first sensor before impinging on the second. The dashed lines indicate the sensor position used for the quotient determination.*

Figures S6.1 and S6.2 show the signals of both sensors in the detector stack. The data is collected for each LED distance by moving the stack behind the lens on the optical axis. The photocurrent in the nano Ampere regime is measured using a Fourier tranform technique[5].
In order to perform distance measurements in the range between 12 m and 72 m, the first sensor is positioned 76 mm and the second 110 mm from the back of the lens. Calculating the ratio of the first sensor signal to the second sensor at these positions for various LED distances yields the calibration curve depicted in Fig. S6.3. The quotient of the sensor signals is shown to monotonically increase over the entire measurement range, making it possible to assign a single quotient value to any distance up to 72 m.

---

[5] The signal is amplified with a Femto DLPCA and recorded with a Behringer U-Phoria UMC202HD sound card

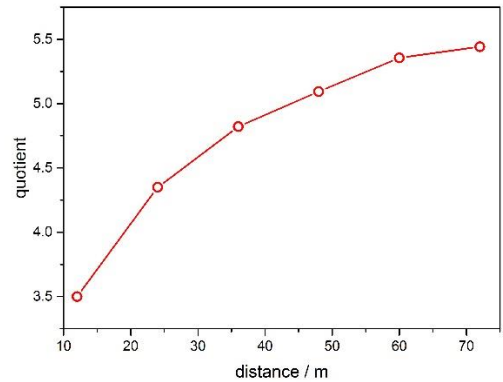

*Fig. S6.3    Calibration curve for distance measurements generated by calculating the quotient of the two sensor signals at 12m, 24 m, 36 m, 48 m, 60 m and 72 m. The line acts as guide to the eye.*